\documentclass[aps,prl,twocolumn]{revtex4-1} 

\usepackage[utf8]{inputenc}
\usepackage{graphicx}
\usepackage{color}
\usepackage{float}
\usepackage{amsmath}
\usepackage{amsthm}
\usepackage{amsfonts}
\usepackage{amsbsy}
\usepackage{amssymb}
\usepackage{amscd}
\usepackage{bm}
\usepackage{xcolor}
\usepackage{hyperref}
\usepackage{subfig}

\usepackage{verbatim}

\begin{document}

\title{Nonuniversal power law spectra in turbulent systems}

\author{V.~Bratanov$^1$, F.~Jenko$^{1,2}$, D.~R.~Hatch$^{1,3}$, and M.~Wilczek$^4$}
\affiliation{$^1$Max-Planck-Institut f\"{u}r Plasmaphysik, EURATOM Association, 85748 Garching, 
Germany\\
$^2$Max-Planck/Princeton Center for Plasma Physics\\
$^3$Institute for Fusion Studies, University of Texas at Austin, Austin TX 78712, USA\\
$^4$Department of Mechanical Engineering, The Johns Hopkins University, 3400 North
Charles Street, Baltimore MD 21218, USA}

\begin{abstract}
\noindent Turbulence is generally associated with universal power law spectra in scale ranges
without significant drive or damping. Although many examples of turbulent systems do not
exhibit such an inertial range, power law spectra may still be observed. As a simple model
for such situations, a modified version of the Kuramoto-Sivashinsky equation is studied.
By means of semi-analytical and numerical studies, one finds power laws with nonuniversal
exponents in the spectral range for which the ratio of nonlinear and linear time scales is
(roughly) scale-independent.
\end{abstract}

\maketitle
{\em Introduction.} Turbulence can generally be described as spatio-temporal chaos in open systems, brought about
by the nonlinear interaction of many degrees of freedom under out-of-equilibrium conditions. As
such, it is ubiquitous in nature and in the laboratory, and represents a fundamental challenge
to theoretical physics. Power law energy spectra constitute one of the most prominent features
of such systems. A first prediction along those lines was provided for three-dimensional Navier-Stokes
turbulence as early as 1941 by Kolmogorov.\cite{Kolmogorov} The typical physical picture is
that power laws emerge on scales where both energy injection and dissipation are negligible,
i.e., in the so-called inertial range. Here, on the basis of dimensional analysis, the value
of the spectral exponent is considered to be determined entirely by the nonlinear energy
transfer rate, implying universality.

Interestingly, there exist numerous examples of turbulent systems which display (simple or
broken) power laws even in the presence of multiscale drive and/or damping. These include,
e.g., flows generated by space-filling fractal square grids \cite{Peinke}, the mesoscale
dynamics in dense bacterial suspensions \cite{PNAS}, and turbulence in astrophysical \cite{ISM}
and laboratory \cite{Goerler} plasmas. At least in the latter case, numerical
studies suggest that the observed power law exponents are not universal,
however.\cite{Goerler} Instead, they appear to depend on the underlying linear physics of the
system. This finding clearly calls for a theoretical understanding that can also help to
interpret and guide experimental as well as numerical investigations.

In a previous investigation \cite{Guercan}, a simple model for density fluctuation spectra in
magnetized laboratory plasmas was proposed which is based on the notion of disparate-scale
interactions between small-scale eddies and large-scale structures like mean or zonal flows,
also taking into account effective linear drive and/or (eddy/Landau) damping. In this
context, universal broken power laws with an exponential cutoff were predicted. In the
present Letter, we consider an alternative scenario. It is shown that one may obtain
nonuniversal power laws in a certain spectral range if the ratio of the relevant nonlinear
and linear time scales is (roughly) scale-independent there.

{\em Modified Kuramoto-Sivashinsky model.}
To enable a semi-analytical treatment, we will employ a modified version of one of the
simplest models for spatio-temporal chaos and turbulence, the Kuramoto-Sivashinsky equation
(KSE), which was originally put forward to describe turbulence in magnetized plasmas \cite{LaQuey,Cohen}, 
chemical reaction-diffusion processes \cite{Kuramoto}, and flame front propagation.\cite{Sivashinsky}
In general, it can be used for the study of nonlinear, spatially extended systems driven far from
thermodynamic equilibrium by long-wavelength instabilities in the presence of appropriate
(translational, parity, and Galilean) symmetries, and subject to short-wavelength damping.
In its one-dimensional form, it reads
\begin{equation}\label{kur_siv}
u_t = -u u_x - \mu u_{xx} - \nu u_{xxxx}
\end{equation}
for the velocity field $u(x,t)$ with the positive parameters $\mu$ and $\nu$. The equation
is supplemented by the periodic boundary condition $u(L,t) = u(0,t)$ for all $t \geq 0$ and
the initial condition $u(x,t=0) = u_{0}(x)$. Considering only functions that belong to
$C^4(\Omega) \cap L^2(\Omega)$ ensures that the system has finite total kinetic energy.
Eq.~(\ref{kur_siv}) can be rewritten in dimensionless units by substituting
$u\rightarrow \mu u/L$, $t \rightarrow tL^2/\mu$, $x\rightarrow Lx$, and
$\nu \rightarrow L^2\mu\nu$. The non-dimensionalized form of the equation is the same as before,
with the modification $\mu = 1$. In the following, we keep the damping parameter $\nu$ undetermined,
but all quantitative results are obtained with $\nu = 1$. The second- and fourth-order spatial
derivatives on the right-hand side of of Eq.~(\ref{kur_siv})
provide an energy source and sink, respectively. Similar to
three-dimensional Navier-Stokes turbulence, energy is injected on large scales and dissipated
on small scales, with the nonlinear term providing the inter-scale transfer.

The periodic boundary conditions suggest a representation of $u(x,t)$ in terms of a Fourier
series defined as
\begin{equation}
u(x,t) = \sum_{n\in\mathbb{Z}}\widehat{u}(k_n,t)e^{ik_n x} \,,
\end{equation}
where the wave numbers $k_n =n\,(2\pi/L)$ are discrete and $n\in\mathbb{Z}$. From the condition
that $u(x,t)$ is real, it follows that $\overline{\widehat{u}(k_n,t)} = \widehat{u}(-k_n,t)$ where
the overbar denotes complex conjugation. Expressing Eq.~(\ref{kur_siv}) in terms of Fourier
coefficients gives
\begin{equation}\label{f_kur_siv}
\widehat{u}_t(k_n) = -\frac{ik_n}{2} \sum_{m\in\mathbb{Z}}
\widehat{u}(k_n-k_m)\widehat{u}(k_m) + (k_n^2 - \nu k_n^4) \widehat{u}(k_n)
\end{equation}
where we have suppressed the time dependence for the ease of notation. Linearly, each mode is
characterized by the drive/damping rate $\gamma=k_n^2 - \nu k_n^4$. The nonlinear term does not
inject or dissipate energy (i.e., summed over $n$, it gives zero), but only redistributes it among
the modes.

\begin{figure}[h!]
  \centering
  \includegraphics[scale=0.25]{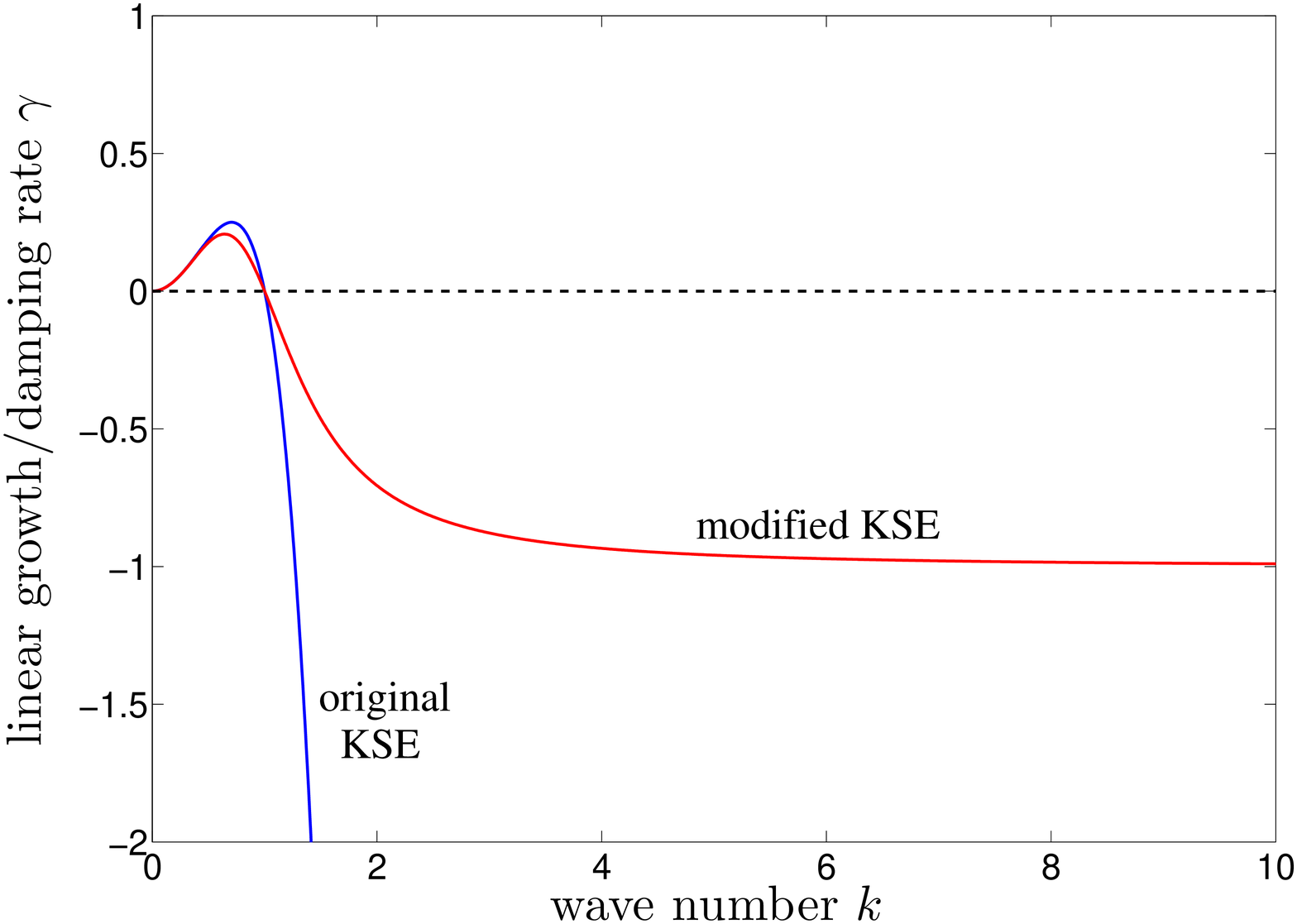}
  \caption{Linear growth/damping rate $\gamma$ as a function of wave number $k$ for the original and
    modified Kuramoto-Sivashinski equation. (color online)}
  \label{KSE-plots}
\end{figure}

We now change the linear term according to
$(k_n^2 - \nu k_n^4)\rightarrow (k_n^2 - \nu k_n^4)/(1 + bk_n^4)$,
such that we obtain the modified KSE
\begin{equation}\label{KS-mod}
\widehat{u}_t(k_n) = -\frac{ik_n}{2} \sum_{m\in\mathbb{Z}}
\widehat{u}(k_n-k_m)\widehat{u}(k_m) + \frac{k_n^2 - \nu k_n^4}{1 + bk_n^4} \widehat{u}(k_n)
\end{equation}
with a constant
damping rate of $\nu/b$ in the high wave number limit (see Fig.~\ref{KSE-plots}). One motivation for
such a modification comes
from the (gyro-)kinetic theory of magnetized plasmas where the growth rates of linear instabilities
tend to a negative constant for large perpendicular wave numbers.\cite{Goerler} Moreover, this is one
of the simplest realizations of a controlled deviation from the classical inertial range. Note that
the real-space representation of the modified linear term is well defined for all functions in the
domain $C^4(\Omega)\cap L^2(\Omega)$.

{\em Energetics.}
The energy budget equation corresponding to the modified version of
Eq.~(\ref{f_kur_siv}) reads in Fourier space
\begin{equation}\label{E_budget}
\frac{\partial E(k_n,t)}{\partial t} = \sum_{m\in\mathbb{Z}}T(k_n,k_m,t) + 
2\frac{k_n^2 - \nu k_n^4}{1+bk_n^4}E(k_n,t)\,,
\end{equation}
where $T(k_n,k_m,t) = k_n\Im(\overline{\widehat{u}(k_n,t)}\widehat{u}(k_n-k_m,t)
\widehat{u}(k_m,t))$ and $E(k_n,t) = |\widehat{u}(k_n,t)|^2$. We shall call the latter
the energy of the $k_n$ mode, while $T(k_n,k_m,t)$ will be referred to as the nonlinear energy
transfer function. In contrast to incompressible fluid turbulence, it is not
antisymmetric with respect to an interchange of $k_n$ and $k_m$. Eq.~(\ref{E_budget}) reflects
the fact that energy transfer takes place via three-wave interactions with $k_n + k_m + k_q = 0$. 
This transfer is conservative, i.e.,
\begin{equation}
\partial_t\mathcal{E}(k_n,t) + \partial_t\mathcal{E}(k_m,t) + \partial_t\mathcal{E}(k_q,t) = 0
\end{equation}
where $\mathcal{E}$ denotes the energy of a mode in a purely nonlinear subsystem that has been
truncated to the three wave numbers $k_n$, $k_m$, and $k_q$. In a quasi-stationary turbulent state,
the time average (denoted by $\langle \cdot \rangle_\tau$) of Eq.~(\ref{E_budget}) reads
\begin{equation}\label{av_E_budget}
\sum_{m\in\mathbb{Z}}\langle T(k_n,k_m,t) \rangle_\tau + 
2\frac{k_n^2 - \nu k_n^4}{1+bk_n^4}E(k_n) = 0\,,
\end{equation}
where $E(k_n)$ denotes $\langle E(k_n,t) \rangle_\tau$.

\begin{figure}[h!]
  \centering
  \includegraphics[scale=0.25]{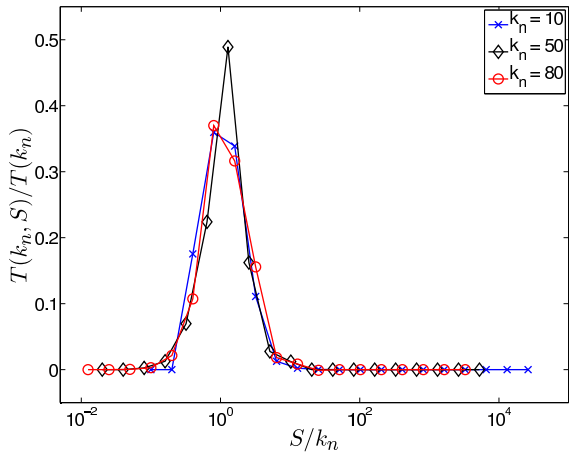}
  \caption{Net energy transfer into mode $k_n$ via triads with the scale disparity parameter
           $S$ as a function of $S$ normalized to $k_n$. (color online)}
  \label{net_transfer}
\end{figure}

{\em Energy transfer physics.}
To gain insight into the turbulent dynamics of Eq.~(\ref{KS-mod}), it is solved numerically,
employing the Exponential Time Differencing fourth-order Runge-Kutta (ETDRK4) algorithm \cite{Cox,Kassam}
and changing the normalized system size to $32\,\pi$.
We focus our investigations on the physics of the net nonlinear energy transfer. As it will turn out,
the latter is dominated by nonlocal interactions in wave number space. Two neighboring high $k$ modes
exchange energy via the coupling to a third mode with $k\sim 1$. This can be quantified by introducing
the scale disparity parameter $S(k,p) = \textrm{max}\{|k|,|p|,|k-p|\}/\textrm{min}\{|k|,|p|,|k-p|\}$
defined in Refs.~\cite{Zhou, NASA}. We shall follow the literature and refer to interactions with small
(large) values of $S$ as local (nonlocal). In Ref.~\cite{NASA}, the observation was made that in Burgers
turbulence, the net energy transfer in the inertial and dissipation ranges is dominated by local
interactions, similar to Navier-Stokes turbulence. Our numerical simulations show that this type
of behavior carries over to the original KSE. To our knowledge, this has not been shown
before. The modified KSE exhibits a completely different scenario,
however. The function $T(k_n,S)$, characterizing the energy transfer into mode $k_n$ via triads with
the scale disparity parameter $S$ and defined over logarithmic $S$-bands like in Ref.~\cite{NASA},
is displayed in Fig.~\ref{net_transfer} as a function of $S/k_n$ for three different values of
$k_n$. In all three cases, one finds a strong peak at $S/k_n\sim 1$, implying that for
Eq.~(\ref{KS-mod}), the net energy transfer at large wave numbers is dominated by nonlocal
interactions, with a $k\sim 1$ mode acting as kind of a catalyst. Nevertheless, the energy
cascade itself is local.
The relevant triadic
interactions can be realized in two different ways: $k_m \approx k_n$ and $k_n-k_m$ small or
$k_m$ small and $k_n-k_m \approx k_n$. Defining for convenience $k_q = k_m - k_n$, the
nonlinearity becomes $k_n\sum_{q\in\mathbb{Z}} P(k_n,k_q)$ where the summand represents the
triple correlation $\Im(\langle \overline{\widehat{u}(k_n,t)}\overline{\widehat{u}(k_q,t)}
\widehat{u}(k_n+k_q,t) \rangle_\tau)$. Considering the numerical results mentioned before, we
have the following picture of the energy transfer in Fourier space. A large mode $k_n$
receives energy (on average) mainly from the mode $k_n - k_d$, where $k_d$ is a relatively small
wave number in the drive range that mediates the transfer. Part of this energy is dissipated
and the rest is forwarded primarily to the mode $k_n + k_d$ again via $k_d$. The first term in
Eq.~(\ref{av_E_budget}) has to balance the energy dissipated by the $k_n$ mode which is the
difference between the energy received by $k_n$ and the one given by $k_n$.
\begin{figure}[h!]
  \centering
  \includegraphics[scale=0.25]{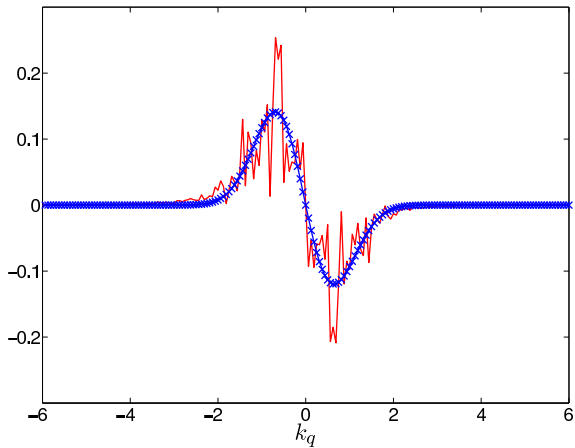}
  \caption{Triple correlation normalized to $E(k_n+k_q^{min})$ (red line) as a function of
           $k_q$ for $k_n = 50$ and $b = 0.036$ compared to the model
           $f_P(k_q)/E(k_n+k_q^{min})$ denoted by blue crosses. (color online)}
  \label{model_f_q}
\end{figure}

{\em Closure model and resulting energy spectra.}
To find a closure model for Eq.~(\ref{av_E_budget}), we search for an approximation of
$P(k_n,k_q)$ at large wave numbers.
The form of $P$ produced by direct numerical simulations is shown in Fig.~\ref{model_f_q}. It
confirms the above picture of nonlinear energy transfer. The
most dominant coupling is indeed with modes in the drive range, and from the minimum and
maximum of the curve one sees that $k_d \approx 1/\sqrt{2}$ which is nearly the linearly most
unstable mode for small $b$. The curve is approximately antisymmetric about $k_q = 0$.
However, it is important that the antisymmetry is not exact: the maximum (at $k_q^{max}
\approx -k_d$) is slightly higher than the absolute value of the minimum (at $k_q^{min}
\approx -k_q^{max}$). This discrepancy is the reason that, at high wave numbers, the spectrum
decreases when $k_n$ increases.
Hence, summing over $k_q$ will lead to a positive contribution that cancels the linear term in
Eq.~(\ref{av_E_budget}) which is negative for high $k_n$. For an approximation of the triple
correlation function $P$ we model the form of the curve in Fig.~\ref{model_f_q} by $f_P(k_q)
= -k_qE(k_n+k_d)\psi_{\xi}(k_q) - k_qE(k_n-k_d)\psi_{-\xi}(k_q)$ where $\psi_{\xi}(k_q)$ is a
localized function centered at and symmetric around $k_q = \xi$ where the value of $\xi$ can
depend on $k_d$. The small asymmetry of $f_P$ is provided by the slightly different prefactors
$E(k_n-k_d)$ and $E(k_n+k_d)$ and $k_q$ ensures the change in sign.

This model allows for an analytically tractable closure of the spectral energy budget
equation at high $k_n$ as
\begin{multline}
\sum_{q\in\mathbb{Z}}f_P(k_q)\approx\frac{1}{\Delta k}\intop_{-\infty}^{+\infty} f_P(k_q)dk_q
=\\ = -\frac{\Phi(\xi)}{\Delta k}\left( E(k_n+k_d) - E(k_n-k_d) \right) \,,
\end{multline}
where $\Phi(\xi) = \int k_q\psi_{\xi}(k_q)dk_q = -\Phi(-\xi)$. Considering that $k_d \approx
1/\sqrt{2} \ll k_n$ we have $E(k_n-k_d) - E(k_n+k_d) \approx -\sqrt{2} dE/dk$ where a
continuum of wave numbers is assumed. Hence,
\begin{equation}
-\frac{1}{\lambda}k\frac{dE}{dk} + 2\frac{k^2 - \nu k^4}{1 + bk^4}E(k) = 0 \,,
\end{equation}
where $\lambda = \Delta k/(2\sqrt{2}\Phi(\xi))$. In physical units, $\lambda$ has the
dimension of time, and at high $k$, $1/\lambda$ can be interpreted as a typical nonlinear
frequency. The factor $2$ takes into account that for
high $k_n$, the nonlinear energy transfer function shows the same structure also at small $k_m$
and large $k_n - k_m$ as we discussed previously. The solution of the above differential
equation is readily obtained as
\begin{equation}
E(k) = \widetilde{E}_0 \exp\left( \frac{\lambda}{\sqrt{b}}\arctan (\sqrt{b}k^2) - 
\frac{\lambda\nu}{2b}\ln (1+bk^4) \right)
\end{equation}
with $\widetilde{E}_0$ a constant of integration. In the limit of large wave numbers the
second term in the exponent dominates and leads to
\begin{equation}\label{power_law}
E(k) = E_0\,k^{-2\lambda\nu /b} \,,
\end{equation}
where $E_0$ is a constant. This is a power law spectrum with a nonuniversal scaling exponent.
The latter is set by the ratio of the linear damping rate $\nu/b$ and the nonlinear frequency
$1/\lambda$.\\
An analytically convenient form for $\psi$ is $\psi(k_q)=a_1 e^{(k_q-a_2)^2/a_3}$ where $a_1$,
$a_2$ and $a_3$ are free parameters. Their values may be determined by a fit to the numerical
data which is shown with blue crosses in Fig.~\ref{model_f_q}. One can easily check that this
particular choice for $\psi$ gives for the ratio between the maximum and the absolute value
of the minimum
\begin{equation}
\frac{f_P(-k_d)}{|f_P(k_d)|} \approx \frac{(1+r e^{4k_da_2/a_3})}
{(1+re^{-4k_da_2/a_3})}e^{-4k_da_2/a_3} \,,
\end{equation}
where $r = E(k_n-k_d)/E(k_n+k_d)$. A least squares fit gives $a_1 \approx 0.1403$, $a_2
\approx 0.2578$ and $a_3 \approx 0.7564$ which leads to $f_P(-k_d)/|f_P(k_d)| \approx 1.217$.
The corresponding numerical value is $1.184$ and the good agreement signifies that the
particular form of $f_{P}$ chosen captures well the important asymmetry of the triple
correlation.
\begin{figure}[ht]
  \centering
  \includegraphics[scale=0.25]{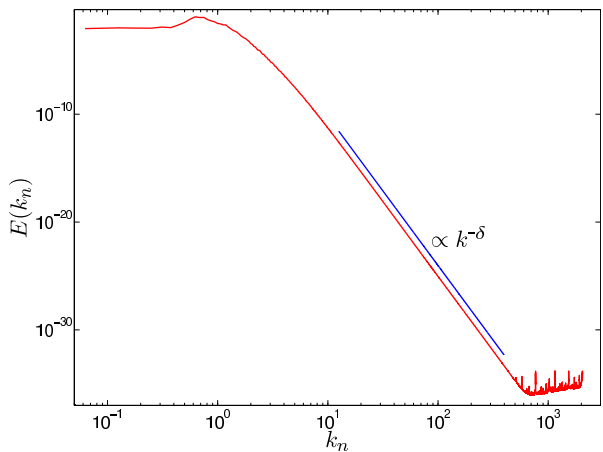}
  \caption{Fit of a power law (blue line) to the high-$k$ end of the energy spectrum (red
           line) for $b = 0.036$. (color online)}
  \label{energy_slope}
\end{figure}

\begin{figure}[ht]
  \centering
  \includegraphics[scale=0.24]{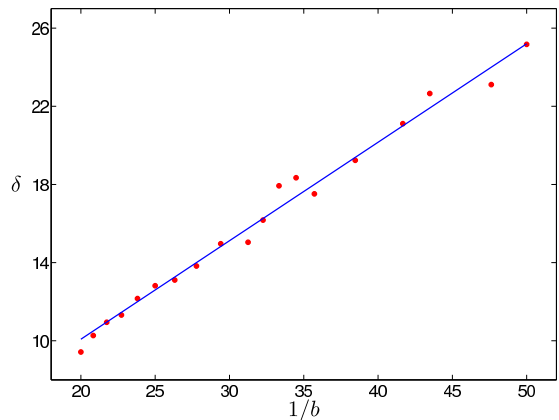}
  \caption{The exponent $\delta=2\lambda\nu /b$ in the power law $E(k)\propto k^{-\delta}$ as
           a function of the damping rate $1/b$ for $\nu=1$. (color online)}
  \label{delta_vs_b}
\end{figure}

{\em Consistency checks.}
To check for consistency, we also computed numerically the energy spectra for different values
of the damping rate $\nu/b$. As can be seen in Fig.~\ref{energy_slope}, one can thus confirm
that a constant high-$k$ damping rate leads to an energy spectrum in the form of a power law
(in contrast to the standard KSE, which displays an exponential fall-off), and that the
associated spectral exponents are indeed proportional to the damping rate. According to a linear
fit to the data in Fig.~\ref{delta_vs_b}, one obtains $\lambda\approx 0.25$, whereas the
fitting procedure in the context of Fig.~\ref{model_f_q} yields a slightly larger value of
$\lambda \approx 0.4$. The reason for this is that the area enclosed by the ragged curve
(which is essential for computing the precise value of the energy transfer) is nearly $1.6$
times larger than the area under the blue curve in Fig.~\ref{model_f_q}. Taking this
correction into account, the two approaches agree very well, providing a consistent overall
picture.

{\em Conclusions.} Motivated by the fact that many turbulent systems in nature as well as in the
laboratory exhibit power law spectra even in the absence of a clean inertial range, we studied
as a simple model system a modified version of the Kuramoto-Sivashinsky equation, with a constant
high-$k$ damping rate. Via semi-analytical and numerical studies, we demonstrated the
existence of power laws with nonuniversal scaling exponents in the spectral range for which
the ratio of nonlinear and linear time scales is (roughly) scale-independent. Such situations
may arise in various physical systems with multiscale drive and/or damping, including, in
particular, magnetized laboratory plasmas.\cite{Teaca} In this context, the present work provides
a plausible explanation for the observation of nonuniversal power laws in numerical
studies.\cite{Goerler}

Another possible application of these findings is kinetic Alfv\'en wave (KAW) turbulence, as it is
thought to occur, e.g., in the solar wind. In this case, one has to compare the nonlinear energy
transfer rates (which scale like $k_\perp^{4/3}$ at sub-ion-gyroradius scales) with the Landau damping
rates of KAWs. The latter may have rather complex $k$ dependencies, with details depending on the
ion-to-electron temperature ratio and the plasma $\beta$.\cite{Howes} There seem to exist parameter
regimes and $k$ ranges for which the ratio of linear and nonlinear frequencies is roughly
scale-independent, such that nonuniversal power-law spectra may emerge.

\section*{Acknowledgements}
We would like to gratefully acknowledge fruitful discussions with A.~Ba\~n\'on Navarro and T.~G\"orler.
The research leading to these results has received funding from the European Research Council under the
European Union's Seventh Framework Programme (FP7/2007-2013) / ERC grant agreement no. 277870.


\begin{thebibliography}{100}

\bibitem{Kolmogorov}
A.~N.~Kolmogorov,
Dokl.~Akad.~Nauk SSSR 30, 299-303 (1941)

\bibitem{Peinke}
R.~Stresing, J.~Peinke, R.~E.~Seoud, and J.~C.~Vassilicos,
Phys.~Rev.~Lett.~104, 194501 (2010)

\bibitem{PNAS}
H.~H.~Wensink {\em et al.},
Proc.~Natl.~Acad.~Sci.~U.S.A.~109, 308 (2012)

\bibitem{ISM}
B.~G.~Elmegreen and J.~Scalo,
Annu.~Rev.~Astron.~Astrophys.~42, 211 (2004)

\bibitem{Goerler}
T.~G\"orler and F.~Jenko,
Phys.~Rev.~Lett.~100, 185002 (2008); Phys.~Plasmas 15, 102508 (2008)

\bibitem{Guercan}
\"O.~D.~G\"urcan {\em et al.},
Phys.~Rev.~Lett.~102, 255002 (2009)

\bibitem{LaQuey}
R.~E.~LaQuey, S.~M.~Mahajan, P.~H.~Rutherford, and W.~M.~Tang,
Phys.~Rev.~Lett.~34, 391 (1975)

\bibitem{Cohen}
B.~I.~Cohen, J.~A.~Krommes, W.~M.~Tang, and M.~N.~Rosenbluth,
Nucl.~Fusion~16, 971 (1976)

\bibitem{Kuramoto}
Y.~Kuramoto and T.~Tsusuki,
Prog.~Theor.~Phys.~52, 1399 (1974); Prog.~Theor.~Phys.~Suppl.~64, 346 (1978)

\bibitem{Sivashinsky}
G.~I.~Sivashinsky,
Acta Astron.~4, 1177 (1977); 6, 560 (1979)

\bibitem{Cox}
S.~M.~Cox, P.~C.~Matthews,
J.~Comp.~Phys.~176, 430 (2002)

\bibitem{Kassam}
A.-K.~Kassam, L.~N.~Trefethen,
SIAM J.~Sci.~Comput.~26, 1214 (2005)

\bibitem{Zhou}
Y.~Zhou,
Phys.~Fluids A 5, 2511 (1993)

\bibitem{NASA}
S.~S.~Girimaji and Y.~Zhou,
Phys.~Lett.~A 202, 279 (1995)

\bibitem{Teaca}
B.~Teaca {\em et al.},
Phys.~Rev.~Lett.~109, 235003 (2012)

\bibitem{Howes}
G.~G.~Howes {\em et al.},
Astrophys.~J.~651, 590 (2006)


\end{thebibliography}
\end{document}